\documentclass[9pt]{article}

\usepackage[margin=1in]{geometry}
\usepackage{cite}
\usepackage{amsmath,amssymb,amsfonts}
\usepackage{algorithmic}
\usepackage{graphicx}
\usepackage{textcomp}
\usepackage{comment}
\usepackage{subcaption}
\usepackage{url}
\usepackage{multirow}
\usepackage{float}

\begin{document}%
\title{Stimulated Secondary Emission of Single Photon Avalanche Diodes }
\date{February 14th, 2024}
\author{\shortstack{Kurtis Raymond, Fabrice Retière, Harry Lewis, Andrea Capra, Duncan McCarthy,\\ Austin de St Croix, Giacomo Gallina,  Joe McLaughlin, Juliette Martin, Nicolas Massacret,\\  Paolo Agnes, Ryan Underwood, Seraphim Koulosousas, Peter Margetak }
\thanks{K. Raymond is with TRIUMF, 4004 Wesbrook Mall, Vancouver, BC V6T 2A3, Canada (e-mail: kraymond@triumf.ca).}
\thanks{F. Retière, A. Capra, N. Massacret, R. Underwood, H. Lewis, and A. de St Croix, are with TRIUMF, 4004 Wesbrook Mall, Vancouver, BC V6T 2A3, Canada}
\thanks{G.~Gallina is with Physics Department, Princeton University, Princeton, NJ 08544, USA}
\thanks{D. McCarthy is with TRIUMF, 4004 Wesbrook Mall, Vancouver, BC V6T 2A3, Canada and UBC, Vancouver, BC V6T 1Z4, Canada}
\thanks{Seraphim Koulosousas are with Royal Holloway, University of London Egham Hill, Egham TW20 0EX, United Kingdom}
\thanks{Paolo Agnes is with GSSI, viale francesco Crispi 7, 67100, L'Aquila, IT, and was supported by Marie Curie programme funding, grant 101029822}}

\maketitle

\begin{abstract}
Large-area next-generation physics experiments rely on using Silicon Photo-Multiplier (SiPM) devices to detect single photons, which trigger charge avalanches. The noise mechanism of external cross-talk occurs when secondary photons produced during a charge avalanche escape from an SiPM and trigger other devices within a detector system. This work presents measured spectra of the secondary photons emitted from the Hamamatsu VUV4 and Fondazione Bruno Kessler VUV-HD3 SiPMs stimulated by laser light, near operational voltages. 
The work describes the Microscope for the Injection and Emission of Light (MIEL) setup, which is an experimental apparatus constructed for this purpose. Measurements have been performed at a range of over-voltage values and temperatures from 86~K to 293~K. 
The number of photons produced per avalanche at the source are calculated from the measured spectra and determined to be 49$\pm$10 and 61$\pm$11 photons produced per avalanche for the VUV4 and VUV-HD3 respectively at 4 volts over-voltage. No significant temperature dependence is observed within the measurement uncertainties. The overall number of photons emitted per avalanche from each SiPM device are also reported.

\end{abstract}

\section{Introduction}
\label{sec:introduction}

Silicon Photomultipliers (SiPMs) have emerged as a compelling solution for detecting single photons in a wide range of applications including particle physics and medical imaging~\cite{Acerbi2018}. They have sub-nanosecond timing resolution in addition to being compact and insensitive to magnetic fields~\cite{sun_study_2018}. In this paper, we focus on the characterization of SiPMs developed by Fondazione Bruno Kessler (FBK), the VUV-HD3, and Hamamatsu Photonics (HPK), the VUV4, for use in liquid Xenon in the context of the nEXO experiment \cite{adhikari_nexo_2021}.

The most relevant features of these devices are their sensitivity in the vacuum ultraviolet range (higher than 15\% efficiency at 175~nm for $\geq$ 2~V over-voltage) and very low dark noise rate (less than 1~Hz/mm$^2$ at 3~V over-voltage) at liquid Xenon temperature (about 168~K)~\cite{gallina_performance_2022}.

A SiPM is an array of single photon avalanche diodes (SPADs), that are electrically isolated from each other and operated above the device-dependent breakdown voltage, $V_\text{br}$.
Incident photons on the device surface will produce a charge ``avalanche," corresponding to a single SPAD's full discharge.
The charge of an avalanche is given by the product of the diode capacitance, $C_d$, and the over-voltage $V_\text{oV} = V - V_\text{br}$, where $V$ is the device operating voltage.
Photon counting is achieved in SiPMs by counting the number of avalanches.
However, the avalanche process produces further secondary photons, resulting in internal and external cross-talk noise mechanisms.

Internal cross-talk refers to secondary photons that trigger avalanches in neighbouring SPADs of the same SiPM without escaping from the SiPM itself. This includes secondary photons that escape from the surface of one SPAD, reflect back into the SiPM at the surface coating interface, and trigger avalanches in neighbouring SPADs \cite{gundacker2020silicon}. This process has been extensively studied in literature with both measurements \cite{Otte2016} and simulation \cite{NepomukOtte2009}. For the two devices of interest in this work, internal cross-talk was found to be more than 20\% for the FBK SiPM but less than 5\% for the HPK device at 3~$V_\text{ov}$~\cite{gallina_performance_2022}.

External cross-talk is the the process that lead to avalanches in separate SiPMs. This process can be problematic for large surface area, SiPM-based detectors such as nEXO. In this configuration each SiPM can trigger other SiPMs in their vicinity, thus contributing to the detector background. It is important to study the SiPM secondary photon emission to quantify the systematic effects hindering the overall detector performance.

This paper's scope is the characterization of light production in SiPM avalanches by recording the number of photons that escape the SiPMs and their wavelengths. This work is to help develop models of photon production in SPADs, and to predict the impact of external cross-talk on future experiments when embedded within simulations which model photon transport and detection efficiency. 

This work builds upon earlier studies characterizing the light produced by thermal avalanches \cite{mclaughlin_characterisation_2021}.
However, the measurements were performed operating the SiPMs at very high over-voltage in order to gather sufficient light.
SiPMs are not used under such conditions in normal operation.  This new study uses a focused laser beam to overcome the lack of photon counts at low over-voltages by forcing avalanches within a specific SPAD, therefore increasing the avalanche rate within the field of view by several orders of magnitude.
This strategy enables reducing systematic errors, facilitating comparison with earlier work \cite{newman_visible_1955, chynoweth_ag_photon_1956, lacaita_bremsstrahlung_1993, mirzoyan_light_2009} (albeit with different devices), and eventual comparison with theoretical models \cite{bude_hot-carrier_1992,villa_photon_1995, akil_multimechanism_1999}.

\section{Experimental setup and technique}

\subsection{Apparatus: the Microscope for the Injection and Emission of Light}

\begin{figure}
    \centering
    \includegraphics[height=6in]{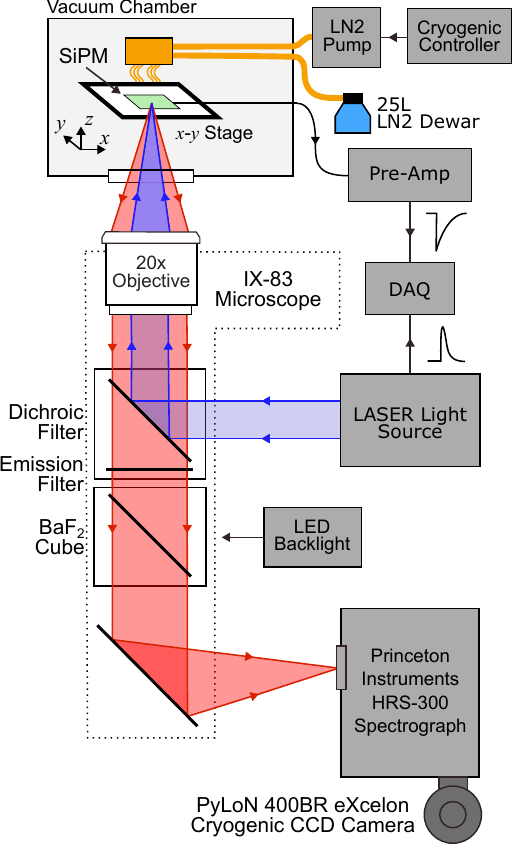}
    \caption{Illustration of the optical path diagram of MIEL experimental setup used in this work. The injected light is labeled as blue, while the re-emitted light is labelled as red.}
    \label{fig:MIEL Optical Path}
\end{figure}

This paper relies on a setup, called the Microscope for the Injection and Emission of Light (MIEL), which is capable of characterizing stimulated emission from single SPADs. MIEL is a confocal microscope including a cryogenic stage as shown in Figure \ref{fig:MIEL Optical Path}.
It is built around an Olympus IX-83 inverted microscope. 
A vacuum chamber is affixed above the microscope with a broadband anti-reflective coated sapphire window separating the vacuum space from the air space within the microscope.
Inside the vacuum chamber is a Micronix sub-micron cryogenic $x$-$y$-positioning stage coupled to a cold finger through copper braided conduction cables.
The cold finger is cooled using a liquid Nitrogen (LN2) suction system, which consists of a 25L LN2 Dewar and an Instec LN2-P suction pump.
Temperature control is achieved with an Instec MK1000 controller, which interfaces with an instrument PC, and is capable of sub 0.1K stability.
The temperature of the SiPM can be set from room temperature to 86K.
The system requires 6-8~hrs to reach a stable temperature to prevent micron-scale positional stage or focus drifts.
In this work, both FBK VUV-HD3 and HPK VUV4 devices are characterized at temperatures ranging from 293~K to 86~K, to match the temperatures of LXe and LAr detectors.

\begin{figure}
    \hfill
    \begin{subfigure}{0.45\columnwidth}
    \caption{Reflected Light}
    \centering
    \includegraphics[width=2.6in]{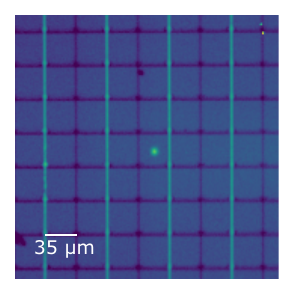}
    \end{subfigure} 
    \hfill
    \begin{subfigure}{0.45\columnwidth}
    \centering
    \caption{Emitted Light}
    \includegraphics[width=2.6in]{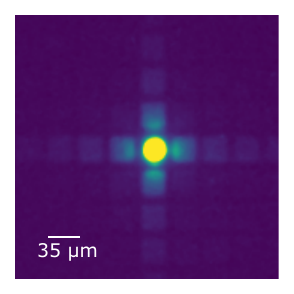}
    \end{subfigure}
    \hfill

    \caption{False colour images (a) show the MIEL setup inspecting the FBK-VUV-HD3 device with the LED backlight on. The spot in the centre of a SPAD is from the injection of light from a broadband halogen lamp. The halogen lamp is replaced with a pulsed stimulating laser during data taking. Image (b) shows the LED backlight off, with the dichroic filter engaged and stimulated using the PicoQuant 405~nm laser source. The formed pattern is likely due to optical cross-talk from the stimulated SPAD.}
    \label{fig:SiPMImage}
\end{figure}

The optical path is as follows:
\begin{itemize}
    \item The stimulation light source for this work is a 405~nm PicoQuant laser head, controlled by a PDL 800-D controller. The laser head is fiber coupled to an OZ Optics variable digital attenuator. The laser was usually operated at a moderate power setting to ensure stable operation and further attenuated to generate an avalanche with 50 - 80\% of laser pulses. The repetition rate ($f_\text{laser}$) was set to 250~kHz to allow the SPAD to fully recharge between laser pulses, while maximizing the number of avalanches per unit time.
    \item The laser light is injected into the microscope using a Thorlabs RC12FC-P01 off-axis parabolic mirror and is optionally further collimated using a manual adjustable aperture ring.
    \item The laser light is reflected into the objective by a 445~nm long-pass dichroic mirror. The objective used for this study is an Olympus LCPLN20XIR with a magnification of 20x and a Numerical Aperture of 0.45. A picture of the laser spot reflected by the SiPM is shown in Figure~\ref{fig:SiPMImage}a. When taking such pictures, an LED illumination source can be added, and is not used in normal operation.
    \item Most of the laser light is absorbed by the SiPMs. The reflected light is eliminated by the dichroic mirror and by an additional 550~nm long pass filter.

    \item The SiPM light is transmitted to a Princeton Instruments HRS 300 Spectrograph coupled to a Princeton Instruments PyLoN 400BR eXcelon cryogenic camera with a backside illuminated CCD sensor. The camera is specially designed for spectroscopic applications, featuring a sensor with an aspect ratio of 1:3.35 and an array of 1340x400 20$\times$20$~\mu$m pixels.  Figure~\ref{fig:SiPMImage}(b) shows an image of the emitted light as measured by the PYLON camera with the spectrograph set to zeroth order, i.e. reflecting all wavelengths. In spectroscopy mode, the 400 vertical pixels are mapped to the physical location of the SPAD, and the 1340 pixels are mapped to a calibrated set of wavelengths. The width of the spectrograph entrance slit was chosen to make the entire active-area of a single SPAD visible. This leads to slit-widths of 35-50~um. The spectral resolution is greatly reduced, but the spectra structure is not expected to have fine features. Photons between 550-1050~nm only are considered due to the lack of photons of wavelengths less than 550~nm escaping the silicon, and the lack of photons interacting with silicon above 1050~nm.
\end{itemize}

For this work, the spectrograph was configured using a 150~g/mm grating with 800~nm blaze, but the system is also equipped with another diffraction grating for visible light.
The setup is capable of allowing light to pass through zeroth order (imaging mode) or at some centre wavelength (Spectroscopy Mode).
The PyLoN camera was operated at 163~K to limit thermal noise and allow for longer exposures.
Reducing the operating camera temperature reduces the efficiency at higher wavelengths, which is be accounted for in the transmission estimate.
The wavelength calibration is performed using the Princeton Instrument IntelliCal® standard ($<0.2$~nm RMS error) Neon-Argon Lamp.
Due to its narrow line widths, the alignment of the camera to the HRS-300 spectrograph is checked at the same time (important for summing vertical pixels).

The SiPM is biased with a Precision Power Supply, Keithley 2280S-60-3. The bias supply is connected with a coax cable to the vacuum feedthrough, which is then converted to a triax cable. This system allows for both pulse readout configuration and optional current-voltage measurements. Waveforms were recorded using a CAEN DT730 digitiser during the first $\sim$30~s of an exposure, corresponding to 100,000 waveforms per setting having 5,000 samples with 2~ns per bin. Waveform analysis was performed to assess the probability that the laser pulse triggered an avalanche. The signal from the SiPM and the synchronous signal from the laser were simultaneously digitized.

\subsection{Analysis}

The experimental goal of this paper is to measure the number of photons, $n_\gamma$, produced by SiPMs per avalanche as a function of wavelength, temperature and over-voltage. In this section, we describe the analysis technique to assess $n_\gamma(\lambda)$, which is the number of photons emitted by the SiPMs within the objective acceptance (Numerical Aperture less than 0.45) as function of wavelength $\lambda$. The steps of the analysis chain are as follows:
\begin{itemize}
    \item Calculate the number of photo-electrons from raw signal produced by the PYLON, which leads to the estimation of the number, $N_\gamma^c(\lambda)$, of photo-electrons detected by the camera at a certain wavelength, $\lambda$, and its associated error. The $c$ superscript denotes the fact that only photons emitted by the SPAD at the centre of the objective are selected, which is accomplished by using a slit at the entrance of the spectrometer to select the column of SPADs. The second dimension is selected by summing the pixels (35 or 50) corresponding to the central SPAD of interest.
    \item Correct for the efficiency of the MIEL apparatus, which leads to the estimation of the correction factor $\epsilon(\lambda)$ with associated systematic error.
    \item Avalanche counting. This number, $N_{av}^c$, is the number of avalanches occurring in the centre SPAD during the exposure. 
\end{itemize}    

The various components of the analysis chain are then combined as follows:

\begin{equation}\label{eq:photon_counting_small}
    n_{\gamma}(\lambda) = \frac{ N_{\gamma}^{c}(\lambda)} { \epsilon(\lambda) \cdot N_{av}^{c}}
\end{equation}

The subsequent sections detail each step of the analysis.

\subsubsection{Pylon camera image Processing} \label{sec:Img_proc}

Measurements of SiPM emission are composed of sub-exposures, known as frames, which are averaged together.
Each frame is a 500s exposure.
Multi-frame data allows the rejection of cosmic rays and observation of the X-Y stage stability over time.

The division in frames enables the effective subtraction of the cosmic-ray background.
Cosmic rays produce spatially narrow tracks or blobs over multiple pixels with large amounts of charge deposited, but are randomly distributed over the CCD.
Thus, cosmic rays can be identified by selecting pixels sufficiently different from the same pixels in different frames.
We also assume that the pixels surrounding the main charge signal may collect a small fraction of the charge but are difficult to identify using the above method.
The average charge is then calculated for the 5 frames, ignoring the pixels that are affected by cosmic rays, which means that not all pixels are averaged over the same number of frames.

The raw data of the Princeton Instruments camera is reported in Analog to Digital Units (ADU).
The calibrated ADU to photo-electron gain ($\nu^{e^-}_\text{ADU}=0.65$) is used to convert this data into photo-electrons.
In the wavelength range of interest (500 to 1000nm), the conversion from photo-electron units to photons observed by the microscope is handled by the apparatus efficiency correction, discussed in Section~\ref{sec:ApparatusEfficiencyCorrection}.

All emission measurements are accompanied by a set of 25-30 frames where the SiPM is unbiased but the laser is still firing, known as a SiPM-off exposure.
The same data processing used in emission measurements is applied to the SiPM-off exposure to reject cosmic-rays.
The final emission exposure is subtracted by its corresponding SiPM-off exposure pixel-by-pixel.

The error in the measurements of the number of photons detected stems from three sources: (i) thermal noise, (ii) ADC read noise, and (iii) photon counting statistics.
The per-pixel dark noise rate of the PYLON camera (0.1789~$e^-$/pixel/hr at -120$^\circ$C) results in 0.075~$e^-$/pixel in a 500~s exposure, which is negligible compared to the read noise of the camera (3.26~$e^-$ RMS when operated at a readout rate of 50~kHz).
Thus only the read noise and counting statistics errors are added in quadrature per pixel.

\subsubsection{Apparatus Efficiency Correction}
\label{sec:ApparatusEfficiencyCorrection}

Sources of apparatus inefficiency include the long optical chain, microscope objectives, mirrors, filters, grating, and camera efficiency. The absolute efficiency is assessed at only a few wavelengths, while the wavelength dependence is measured using an OceanInsight Radiometric Calibrated Light Source HL-3, using the calibration curve provided by the manufacturer. The 400$\mu$m fiber from the OceanInsight light source was placed above the microscope objective at approximately the same location as the SiPM sample to obtain the relative efficiency. Two sets of images were taken with the lamp on and off, each being a collection of 25 frames of 6s exposures. The resulting spectra were processed according to Section~\ref{sec:Img_proc}.

Absolute calibration is found by injecting fiber-coupled laser light at a set wavelength.
The laser light is coupled into a single mode (SM) 50~$\mu$m 0.22NA optical fiber, and a Thorlabs S150C measured the power output of that fiber at a 1kHz repetition rate.
The laser power was unchanged throughout the calibration process, and was monitored for 1~hr and 15~min.
The chosen repetition rate and laser power allowed an adequate average power measurement on the Thorlabs power meter ($\sim$1.5~nA, well above the minimum 10~pA resolution), and sufficiently low enough for the camera to take 50~ms exposures without saturation, which corresponds to roughly 50 laser pulses per exposure. 

\begin{figure}[!t]
\centerline{\includegraphics[width=2.6in]{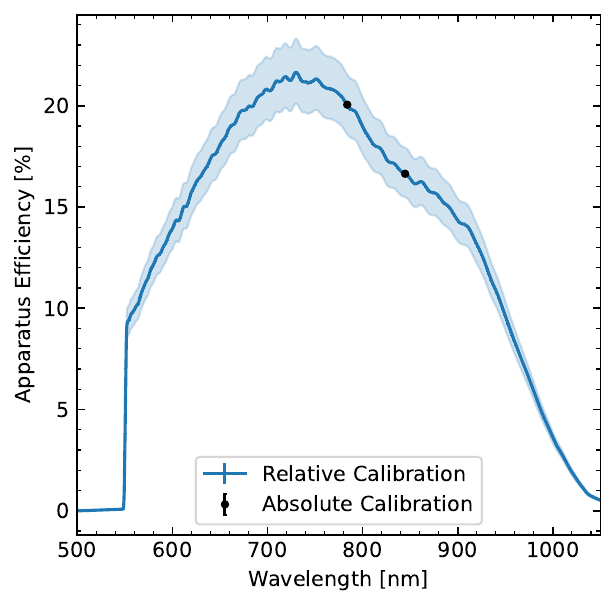}}
\caption{Absolute transmission, $\epsilon(\lambda)$, of the MIEL setup. The relative transmission was found from the calibrated Ocean Optics HL-3, and the absolute transmission found from the method reported in Sec.~\ref{sec:ApparatusEfficiencyCorrection} from a laser source. Systematic bounds derived from the calibration equipment showing the maximum extent of the transmission are shown in light blue. This systematic is common to all measurements made on the MIEL setup.}
\label{fig1}
\end{figure}

The measurement of the fiber with the camera consists of 300 (100 x 3 averaged) exposures.
The ratio of the pulse power calculated from the power meter and the pulse power calculated from the camera results in the transmission, $\epsilon(\lambda)$, at the laser's wavelength.
The maximum systematic uncertainty of performing this measurement is 0.5\% for the Thorlabs PM101, 3\% from the ThorLabs S150C power meter, and 3\% from the HL-3 light source. The efficiency curve is shown in Figure \ref{fig1}. We note that this efficiency is about half what was reported in our earlier paper\cite{mclaughlin_characterisation_2021} because we found that the spectrograph efficiency was lower than expected. 

\begin{figure}[h]
    \centering
    \includegraphics[width=2.6in]{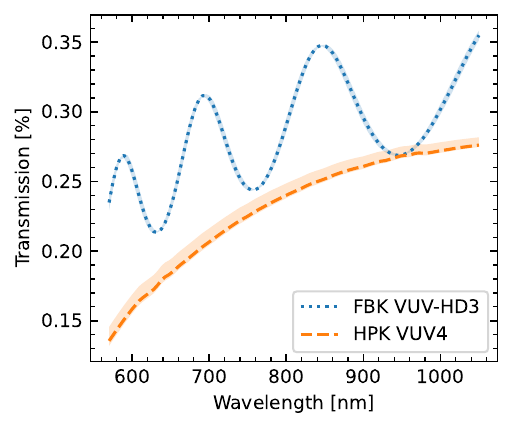}
    \caption{Transmission Calculations from source to (NA=0.45)  microscope objective. The systematic uncertainty is small on the FBK device due to the independent ellipsometer measurement of the thickness of its SiO$_2$ layer. See the text on how the systematic bands are calculated.}
    \label{fig:trans_calc}
\end{figure}

\subsubsection{Avalanche Counting}
\label{sec:Avalanche_Counting}

The extraction of the number of avalanches occurring in the centre SPAD during the camera exposure is performed by analysing waveforms for each setting. The probability that the laser stimulates at least one avalanche, $P_f$, is measured by building an histogram of the maximum amplitude, i.e. the largest value of the waveform with respect to baseline, within a 6 ns window (3 bins), at the time corresponding to the time of the laser induced pulse. This occurs 15 ns after the laser synchronous electrical signal is received. The histogram is only used to count how often the maximum amplitude differs from the baseline, and the number of avalanches per laser flash is not used as it is distorted by internal cross-talk. The number of laser flashes stimulating at least one avalanche during the camera exposure is calculated by multiplying $P_f$ by $R$, the laser repetition rate, and $\Delta t$, the exposure time.

Extracting the number of avalanches in the center SPAD, $N^c_{av}$ requires considering that the laser may produce avalanches in other SPADs than the centre one and that additional delayed avalanches correlated with the laser fired one may occur in the centre SPAD. The former process is parameterized by the probability $P_c$ that a photon from the laser hits the centre SPAD. The latter process is expected due to the after-pulsing process that is known to increase with increasing over-voltage and, to a lesser extent, with decreasing temperature. It is parameterized by the number of correlated delayed avalanches $N_{cda}$ per avalanche. Both processes can be factorized such that:

\begin{equation}
    N^c_{av} = \Delta t \cdot R \cdot P_f \cdot P_c \cdot (1+N_{cda})
\end{equation}

While a quantitative analysis of the laser spot images shows that it is well confined within the centre SPAD, there is a possibility that some of the avalanches triggered by the laser are not within the centre SPAD. Multiple reflection may cause a halo of light that is difficult to image. $P_c$ was determined to be $67.8\pm10.7\%$ by measuring the probability of having exactly zero and one avalanche triggered by the laser at different levels of optical power attenuation. A full description of this calculation is given in appendix \ref{sec:confinement}.

The HPK VUV4 device used in this study is known to exhibit non-negligible after-pulsing \cite{gallina_characterization_2019} and therefore it is expected that a fraction of the avalanches occurring in the centre SPAD are not accounted for when counting avalanches in coincidence with the laser. The nature of our data makes it difficult to estimate $N_{cda}$ in situ. Instead, it is estimated from Ref.~\cite{gallina_characterization_2019} for the HPK VUV4 and Ref.~\cite{gallina_performance_2022} for the FBK VUV-HD3. The VUV4 utilizes CDA values ranging from 233~K to 163~K. The VUV-HD3 device is only corrected using a CDA value near LXe temperatures due to the lack of data published at other temperatures.
The total impact of this correction is on the order of 10\%, especially for the HPK-VUV4 device used in this study, and at higher over-voltages.

Putting all the correction together leads to the following equation as a function of the over-voltage ($\Delta V$) and temperature (T) and for light emitted at the wavelength ($\lambda$):

\begin{equation}\label{eq:photon_counting_full}
    n_{\gamma}(\lambda, \Delta V, T) = \frac{ N_{\gamma}^{cam}(\lambda)} { \epsilon(\lambda) \cdot \Delta t \cdot R \cdot P_{f}(\Delta V, T) \cdot P_c \cdot (1+N_{cda}(\Delta V, T))}
\end{equation}

where $N_{\gamma}^{cam}/P_f$ is measured quantity for each data set. $\Delta t$ and R are operational constants. The $\epsilon$, $P_c$ and $N_{cda}$ variables represent our ability to detect the re-emitted light, focus the laser beam onto one SPAD, and count all the avalanches within one SPAD, respectively. The dominant source of error is the calculation of the setup optical efficiency, $\epsilon$.

\section{Measured light emission spectrum within the apparatus acceptance}

\begin{figure*}[t]
    \centering    
    \captionsetup[subfigure]{justification=centering}
    \begin{subfigure}{0.45\columnwidth}\centering
    \centering  
    \caption{}
    \includegraphics[width=2.6in]{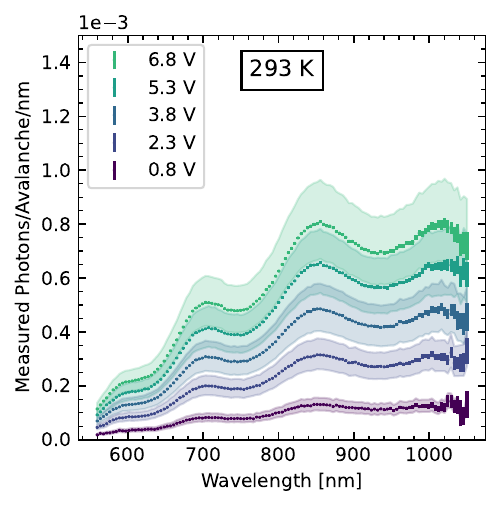}
    \end{subfigure}
    \hfill
    \begin{subfigure}{0.45\columnwidth}\centering
    \centering  
    \caption{}
    \includegraphics[width=2.6in]{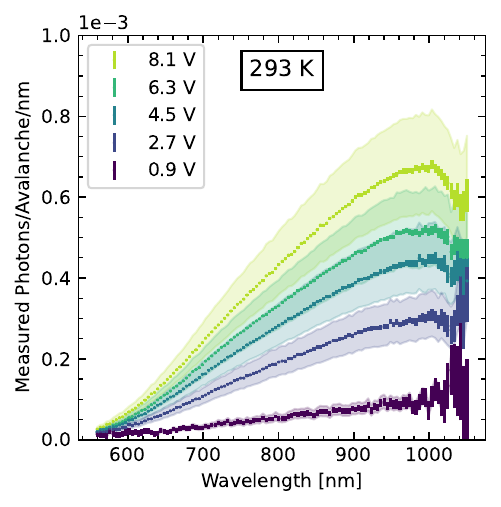}
    \end{subfigure}
    \newline
    \begin{subfigure}{0.45\columnwidth}\centering
    \centering  
    \caption{}
    \includegraphics[width=2.6in]{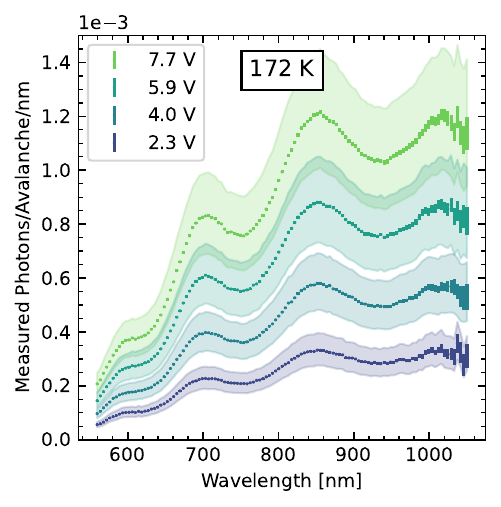}
    \end{subfigure}
    \hfill
    \begin{subfigure}{0.45\columnwidth}\centering
    \centering  
    \caption{}
    \includegraphics[width=2.6in]{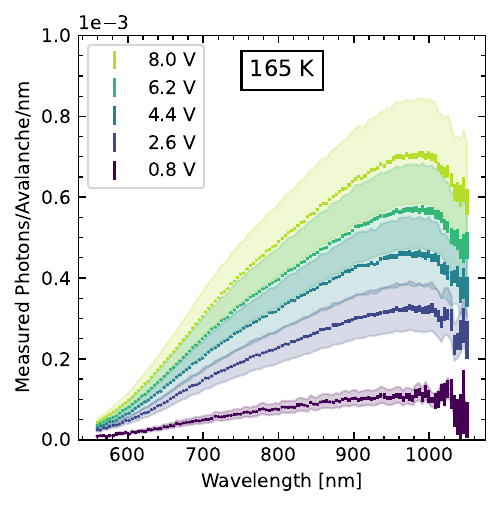}
    \end{subfigure}
    \newline
    \caption{Emission from the FBK-VUV-HD3 (left Column, Figures a and c) and HPK-VUV4 (right Column, Figures b and d) SiPM into a 0.45 NA objective at room temperature (top) and near liquid xenon temperature (bottom). Systematic errors are shown as shaded error bands. Intermediate temperatures have been omitted for clarity.}
    \label{fig:Objective_Emission}
\end{figure*}

Measured emission spectra are shown in Figure \ref{fig:Objective_Emission} for each device as a function of over-voltage and, at room temperature and liquid Xenon temperature. Slightly different cryogenic temperatures were used for each device (172~K for the VUV4 and 165~K for the VUV-HD3) due to limitations in the precision of the set temperature at the measurement stage. The shapes of the measured spectra at intermediate temperatures were similar.

The FBK VUV-HD3 device exhibits oscillations with respect to wavelength, which are observed as a result of interference within the device's micron scale thick SiO$_2$ layer.
This oscillation partially obscure the spectral features of the light emission.
Contrasting this, the HPK-VUV4 device has no oscillations due to interference, and the increasing photon production with wavelength can be observed directly. This is presumably because the SiO$_2$ layer is much thinner. 
In both devices, the maximum emission wavelength is around 1000~nm, above which the number of observed photons decreases rapidly with wavelength.

\begin{figure}
    \centering
    \begin{subfigure}{0.45\columnwidth}\centering
    \caption{}
    \includegraphics[width=2.6in]{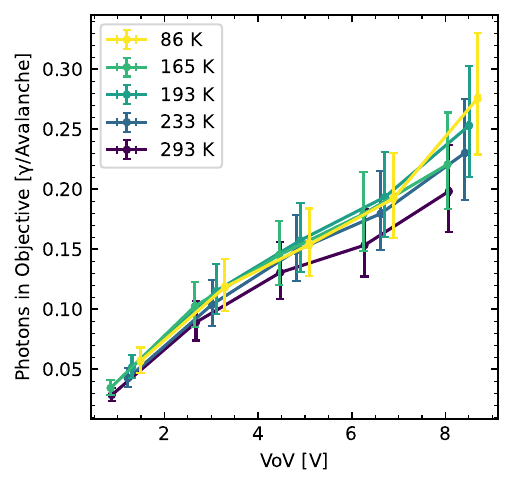}
    \end{subfigure}
    \begin{subfigure}{0.45\columnwidth}\centering
    \caption{}
    \includegraphics[width=2.6in]{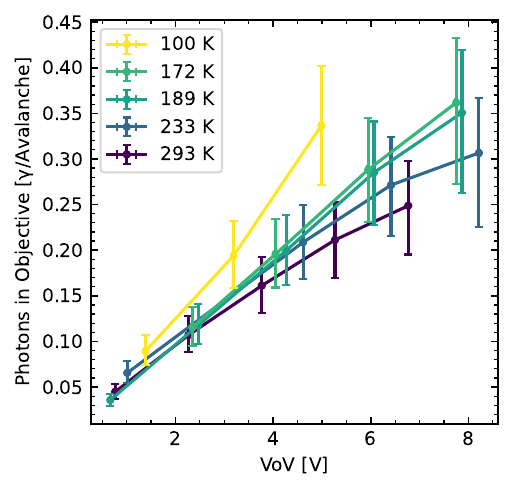}
    \end{subfigure}
     \caption{Photon yield integrated between 550~nm and 1000~nm for the HPK (a) and FBK (b) SiPM into a 0.45 NA objective at different temperatures. Systematic errors are included. The spread of points in the FBK device at high over voltages is due to applying the $N_{cda}$ correction from near LXe temperatures to other temperatures.}
    \label{fig:Objective_Yield}
\end{figure}  

\begin{figure*}
    \captionsetup[subfigure]{justification=centering}
    \begin{subfigure}[b]{0.45\columnwidth}\centering
    \centering
    \caption{}
    \includegraphics[width=2.6in]{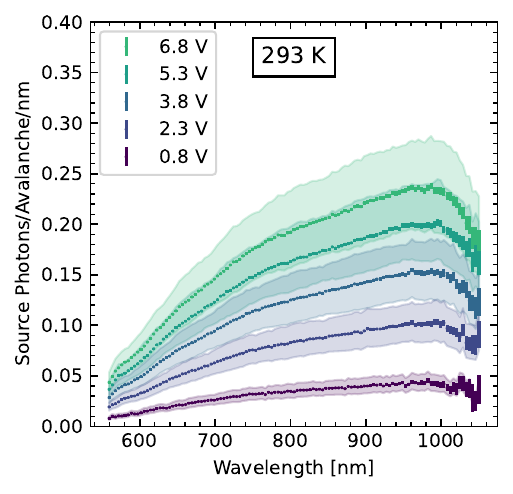}
    \end{subfigure}
    \begin{subfigure}{0.45\columnwidth}\centering
    \centering
    \caption{}
    \includegraphics[width=2.6in]{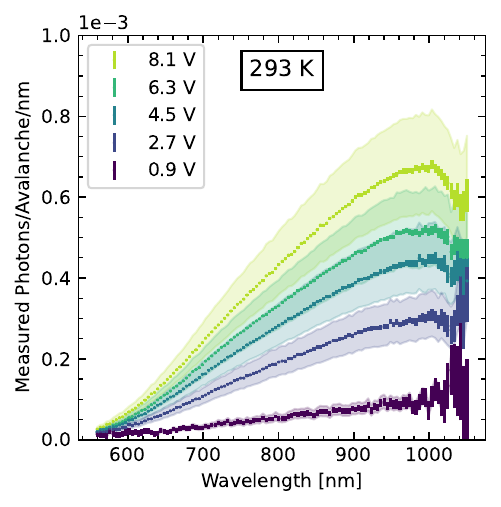}
    \end{subfigure}
    \newline
    \begin{subfigure}{0.45\columnwidth}\centering
    \centering
    \caption{}
    \includegraphics[width=2.6in]{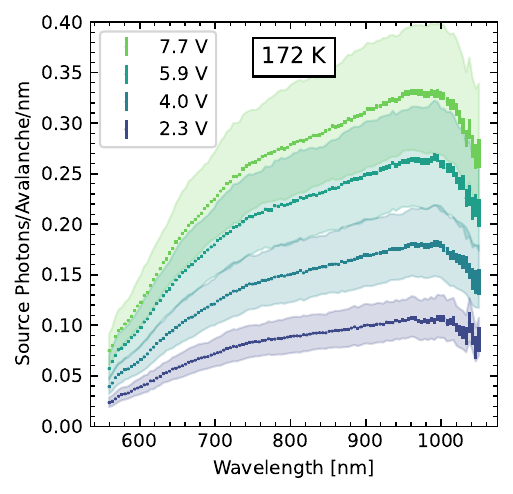}
    \end{subfigure}
    \begin{subfigure}{0.45\columnwidth}\centering
    \centering
    \caption{}
    \includegraphics[width=2.6in]{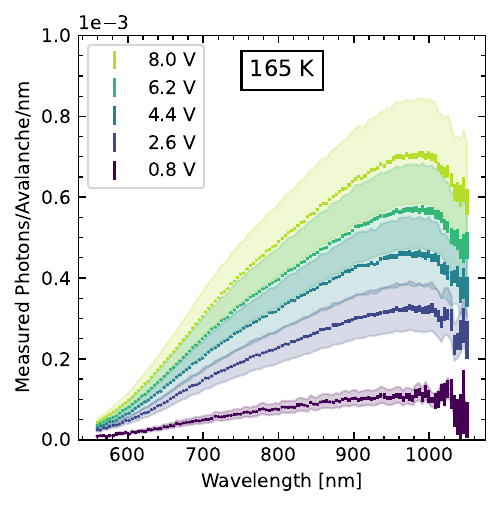}
    \end{subfigure}
    \caption{Average secondary photons generated per avalanche for the FBK VUV-HD3 (left column, Figures a and c) and Hamamatsu VUV4 (right column, Figures b and d) devices, as calculated using the transmission in Fig.~\ref{fig:trans_calc}. Intermediate temperatures have been omitted for clarity.}
    \label{fig:FBK_HPK_Comparions}
\end{figure*}

The yields integrated between 550 nm and 1000 nm are shown in Figure~\ref{fig:Objective_Yield}, including measurements at intermediate temperatures. The yields are in partial disagreement with previous work \cite{mclaughlin_characterisation_2021}.
The latter study used a setup efficiency curve without a direct measurement of the absolute transmission of the system, and instead relied on a model which did not take into account all optical elements within the spectrometer.
This resulted in a higher transmission, and thus lower measured photon counts.

\section{Extrapolation to the source of photons}

In order to extract the number of photons emitted by an avalanche from the spectra measured at the microscope objective, two optical models of the SiPM surface structure were constructed.
Both consist of bulk Silicon with a layer of SiO$_2$ of a specified thickness, $d_{\text{SiO}_2}$ on top, and surrounded by air.
The source of photons is modelled as a point source located a depth $d_p$ within the silicon from the Si-SiO$_2$ interface.
$d_p$ is approximated as the junction depth for each device, and taken as 0.145$\pm 0.01~\mu$m on the VUV-HD3 and 0.8$\pm0.2~\mu$m on the VUV4 \cite{gallina_characterization_2019}. It is assumed that photons are emitted isotropically from the point source.

To constrain the SiO$_2$ thickness for the FBK device, the interference peaks present in the emission spectra were matched to the peaks present in the transmission calculation.
Further refinement was performed by altering the SiO$_2$ thickness in the model until any resemblance of interference in the calculated source spectra were removed.
This method resulted in an SiO$_2$ thickness of 1.4~$\mu$m for this device, which agreed with independent ellipsometry measurements performed on another SiPM from the same wafer batch.

The SiO$_2$ thickness for the HPK device was assumed to be nominally less than 0.1~$\mu$m due to the lack of interference effects observed in the emission.
Using the reflectivity of the HPK-VUV4 device as measured in Ref.~\cite{9246577}, and calculating the reflection using Fresnel reflection equations leads to an SiO$_2$ thickness of $15.9\pm2$~nm.

There are two main contributions to the transmission from the source to objective: (1) attenuation in the silicon and (2) transmission through the Si-SiO$_2$-Air interface.
Let $\theta$ denote the angle of light emitted from a point source at a depth of $d_p$ from the SiO$2$ layer.
The first transmission component is heavily dependent on the attenuation length $\mu(\lambda)$ and can then be written as,

\begin{align}
    T_\text{Si}(\lambda, \theta) = \exp{\left(\frac{-d_p}{\mu(\lambda)\cos \theta }\right)}
\end{align}

Fresnel equations were used to model the multilayered reflections and interference occurring between the Si-SiO$_2$-Air interface \cite{eugene_hecht_optics_2017}, denoted as $T_\text{Si-SiO$_2$-Air}$.
Temperature-dependent refractive indices were not used when performing these calculations.
The refractive indices of Silicon were taken from Refs \cite{rit_center_for_nanolithography_research_optical_2010, schinke_uncertainty_2015}, and LXe from Ref.~\cite{hitachi_new_2005}. 

The transmission vs wavelength can be written as,

\begin{align}\label{eq:SiPM_Transmission}
        A(\lambda) &= \frac{\text{Solid Angle Reaching Objective}}{\text{Solid Angle of the Source}}\\
        &= \frac{\int_{0}^{2\pi}d\phi \int_0^{\theta_{\text{Si}(\lambda)}} T_\text{SiPM}(\lambda, \theta)  \sin(\theta) d\theta}{\int_{0}^{2\pi}d\phi \int_{0}^{\pi}\sin(\theta)d\theta} 
\end{align}

Where,
\begin{equation}
    T_\text{SiPM}(\lambda, \theta) = T_\text{Si}(\lambda, \theta)\cdot T_\text{Si-SiO2-Air} (\lambda, \theta)\label{eq:SiPM_Transmission_total}
\end{equation}

The $\theta_\text{Si}(\lambda)$ parameter can be defined two ways. Objectives are defined by a numerical aperture (NA), which can be used to define $n_\text{Si}\sin\theta_\text{Si}(\lambda) = \text{NA}$.
When considering the total number of photons escaped from the device, i.e. yield, $\theta_\text{Si}(\lambda)$ must be set so that the maximum angle of emission is parallel to the SiO$_2$ layer.
The wavelength dependent transmission calculations are shown in Figure~\ref{fig:trans_calc} 

Figure \ref{fig:FBK_HPK_Comparions} shows both devices as measured and corrected using Eq.~\ref{eq:photon_counting_full} and with the transmission calculations applied.
In the case of the VUV-HD3 device, the oscillations, with respect to wavelength, are entirely accounted for by the optical transmission.
Both emission spectra have a similar shape, with the maximum emission peaking around 1000~nm, before dropping rapidly at higher wavelengths.
The VUV-HD3 device has a small shoulder at 720~nm, whereby the emission does not increase with wavelength as much as below 720~nm.
There is no temperature tend, within the errors, observed on the HPK-VUV4 device, and a slight temperature tend with the FBK-VUV-HD3 device.
However, the latter may be explained by using only the additional correlated avalanche for 173~K and not over a broad range of temperatures.

 It is also of interest to derive the total number of photons which escape the SiPM during an avalanche.
This required an additional calculation using Eq.~\ref{eq:SiPM_Transmission} with the maximum possible angle in Silicon, which is dependent on the media the SiPM is situated in.
This calculation was performed for both air and LXe.
In both media, the FBK device has a higher emission, but both devices have a similar spectral shape.
The results of these calculations are shown in Table \ref{tab:results_table}. The error in determining the transmission is greater when calculating across all angles, resulting in a greater overall systematic error than the calculation of photons produced from a point source within the acceptance of the microscope objective.

\begin{table}
    \centering
    \caption{Summary of calculations and modelling performed with each device. Errors reported are systematic. Values are interpolated to $V_\text{oV}$~=~4V. Capacitance values for each device were taken from Ref.~\cite{gallina_performance_2022}, and each device showed a non-constant trend with over-voltage which may indicate issues with pulse counting at low over-voltages. Other reported values for reference:
    Ref.~\cite{mirzoyan_light_2009}: (500–1117~nm) 1.2$\times10^{-5} \gamma/e^-$, and Ref~\cite{lacaita_bremsstrahlung_1993}: [0.5-4.5]~mA (413-1087~nm) 2.9$\times10^{-5} \gamma/e^-$.
    Values for the photons emitted from the device are modelled in either air ($n=1$) or into LXe (Values for $n$ taken from Ref.~\cite{hitachi_new_2005}).}
        \label{tab:results_table}

    \resizebox{\textwidth}{!}{%
    \begin{tabular}{|c|c||c|l|}
    \hline
    \setlength{\tabcolsep}{3pt}
            \rule{0pt}{8pt}
    \multirow{ 2}{*}{Environmental Conditions}  & \multicolumn{2}{c|}{Device} & \multirow{ 2}{*}{Unit}\\
    \cline{2-3}
    \rule{0pt}{8pt}
    & HPK VUV4 & FBK VUV-HD3 &\\
   \hline
   \hline
       \rule{0pt}{8pt}
    293~K & $0.120\pm0.020$ & $0.169\pm0.032$ & \multirow{ 2}{*}{Photons measured at microscope objective per avalanche} \\
   165~K (VUV4), 172~K (VUV-HD3) & $0.135\pm0.023$ & $0.193\pm0.037$ &  \\
   \hline
       \rule{0pt}{8pt}
 
    \multirow{ 2}{*}{293~K}  & 48.5 $\pm$ 9.5 & 60.8 $\pm$ 10.8 & Source Photons per avalanche \\
      &  (1.94 $\pm$ 0.38)$\times10^{-5}$ & (2.59 $\pm$ 0.46)$\times10^{-5}$ & Source Photons per charge carrier ($e^-$) \\ 
   \hline
       \rule{0pt}{8pt}

     Air: 293~K & 0.64 $\pm$ 0.26 & 0.79 $\pm$ 0.31 & \multirow{ 2}{*}{Photons emitted from SiPM per avalanche}\\
    LXe: 165~K (VUV4), 172~K (VUV-HD3) & 1.55 $\pm$ 0.62 & 1.92 $\pm$ 0.74 &\\
    \hline
    \end{tabular}}
    
\end{table}

\section{Conclusion}

This paper presents a detailed characterization of the light emission process occurring in SiPMs.
Within the acceptance of our objective, we observe $0.17\pm0.03$ (FBK VUV-HD3) and $0.12\pm0.02$ (HPK VUV4) photons emitted per avalanche at $V_\text{oV}=4$~V and at room temperature, which increases with increasing the over-voltage.
Taking into account reflection and absorption of photons produced from their source at the p-n junction to the objective, we estimate that $60.8\pm10.8$ (FBK VUV-HD3) and $49.5\pm9.5$ (HPK VUV4) photons are produced at the source per avalanche near room temperatures.
Both types of SiPMs that we tested are surprisingly close to each other, even though their diode structure is known to differ\cite{gallina_characterization_2019}.
Furthermore, internal cross-talk was found to be 10 times larger for the FBK than for Hamamatsu \cite{gallina_performance_2022}, and internal cross-talk is expected to scale with the number of produced in the avalanche.
This discrepancy can be explained because the VUV4 uses trenches made from very absorbing material (Tungsten) between their SPADs, while the VUV-HD3 trenches are filled dielectric and ineffective at stopping or reflecting photons. 
As shown in Sec.~\ref{tab:results_table}, it is interesting to note that the single SPAD capacitance is similar between the VUV-HD3 and VUV4 devices \cite{gallina_performance_2022}, and therefore the number of electrons produced per avalanches is also of the same order of magnitude in both SiPMs.  
The scale factor is of the order of 1x$10^{-5}$ photons produced per electron produced, which is in good agreement with earlier publications \cite{lacaita_bremsstrahlung_1993, mirzoyan_light_2009}. However, it disagrees with our own measurement made in the dark \cite{mclaughlin_characterisation_2021}. This is explained by the fact that the experimental setup's efficiency was overestimated by a factor of 2 in this earlier publication.
Again, despite large difference in the junction structure for the difference devices tested, the yield per electron is remarkably similar.
The downturn in emission that occurs around 1000~nm on both devices is a striking feature which may provide valuable insight into the emission process in these devices.
Overall, more work is needed to understand whether this is caused by an actual physical process related to the emission process.

Finally, our measurements allow predicting the impact of light emission on experiments using many SiPMs in line of sight of each other, either directly or through reflection.
We provide a measure of the number of photons escaping the SiPMs in air, liquid argon and liquid xenon.
This yield increases with the increasing index of refraction, i.e. it is largest in liquid xenon and lowest in air.
We expect this work to be most valuable in predicting the performance of large detectors searching for rare processes, such as dark matter interactions, in liquid xenon and liquid argon.

\section*{Acknowledgment}

The authors gratefully acknowledge support from Canadian Foundation for Innovation Fund (CFI) 2017. Additional support was provided by a grant from Canada Natural Sciences and Engineering Research Council of Canada for the nEXO project.

\bibliographystyle{ieeetr}
\bibliography{references_fixed}

\section{Determining confinement of laser light to the central SPAD}
\label{sec:confinement}

This section concerns the calculation of $P_c$, which is the probability that a given photon emitted by the laser hits the central SPAD. In order to assess this quantity, the probability of having exactly zero and one avalanche triggered by the laser was measured as a function of the light intensity, varying the optical attenuation parameter $\alpha$ (in dB) such that number of photons hitting the SiPM is N:

\begin{equation}
    N = N_0 \times 10^{-\alpha/10}
\end{equation}

\begin{figure}
    \centering
    \includegraphics[width=0.45\columnwidth]{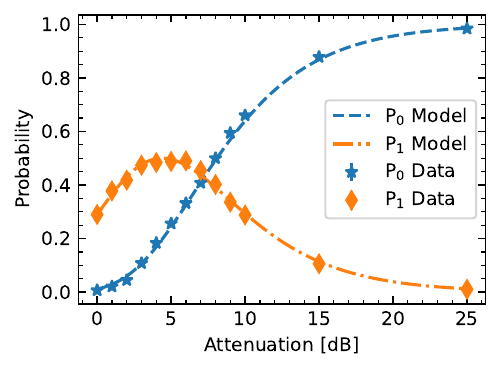}
    \caption{Probability of seeing exactly zero or exactly one avalanches given some amount of attenuation on the laser beam.}
    \label{fig:nPEModelingAttenuation}
\end{figure}

Where $N_0$ is the number of photons hitting the SiPM at 0 dB attenuation. In the absence of any leakage of laser light onto other SPADs, the probability of zero avalanche will converge to zero and the probability of having exactly one avalanche ($P_1$) will converge to one (the centre SPAD always avalanches) times one minus the probability, $P_x$,  of having a least one additional prompt avalanche due internal cross-talk. Figure~\ref{fig:nPEModelingAttenuation} shows however that $P_1$ decreases at sufficiently low value of $\alpha$, i.e. when N is large, which shows that other SPADs are being fired by the laser in addition to the centre SPAD.
$P_0$ and $P_1$ were measured by focusing the laser onto a SPAD of the FBK VUV-HD3 device using the 20x objective, and an oscilloscope was used to histogram the SiPM pulse heights. The oscilloscope was gated on the region immediately after the laser sync signal. $P_0$ and $P_1$ were calculated by counting the number of 0PE pulses and 1PE pulses, and dividing by the total counts in the histogram. It is possible to develop a model to assess $P_c$ by introducing $\epsilon$, the photon detection efficiency at the laser wavelength, and then using binomial statistics for the centre SPAD, and Poisson statistics for the other SPADs as a function of $n$, which denotes the number of photons impinging onto the respective SPAD or group of SPADs. This is shown in Eq. \ref{eq:P0(n)}-\ref{eq:p1(n)}:
\begin{equation}
    p^c_0(n) = (1-\epsilon)^n \\
    \label{eq:P0(n)}
\end{equation}
\begin{equation}
    p^c_1(n) = (1-(1-\epsilon))^n) (1-P_x)\\
\end{equation}
\begin{equation}
    p^o_0(n) = e^{-\epsilon n} \\
\end{equation}
\begin{equation}
    p^o_1(n) = \epsilon n e^{-\epsilon n} (1-P_x)
    \label{eq:p1(n)}
\end{equation}

$p^c_0$ and $p^c_1$ are the probabilities of firing zero and one avalanche if $n$ photons impinge onto the centre SPAD. $p^co_0$ and $p^o_1$ are the probabilities of firing zero and one avalanche if $n$ photons impinge onto any other SPADs. Introducing $n_t$, the total number of photons hitting the SiPM, and $n_c$, the number of photons hitting the central SPAD, the probability of having exactly one avalanche can then be written as:

\begin{equation}
\begin{split}
    p_1(n_t,n_c) = (1-P_x) (p^c_1(n_c)p^o_0(n_t-n_c)+p^c_0(n_c)p^o_1(n_t-n_c)) \\
       = e^{-(n_t-n_c)\epsilon} [(1- (1-\epsilon)^{n_c})+(1-\epsilon)^{n_c} (n_t-n_c)]
\end{split}    
\end{equation}

Then the measured quantities $P_0$ and $P_1$ can be calculated by using Poisson statistics for the total number of photons hitting the SiPM, $n_t$ and using binomial statistics to assess the number of photon hitting the centre SPAD, $n_c$ such that:

\begin{align}
    P_0(N) = \sum_{n_t=1}^{\infty} \text{Pois}(n_t,N) (1-\epsilon)^{n_t} \label{eq:P_0} \\
    P_1(N) = \sum_{n_t=1}^{\infty} \text{Pois}(n_t,N) \sum_{n_c=0}^{n} \text{Bin}(n_c,n,P_c) p_1(n,n_c) \label{eq:P_1}
\end{align}

Both Eq.~\ref{eq:P_0} and Eq.~\ref{eq:P_1} were fit simultaneously to the data, as shown in Figure~\ref{fig:nPEModelingAttenuation}, with $P_c$ being the free parameter and $\epsilon$ being fixed at 0.6.
The fit shows a reasonable trend to the data, and produces a probability of $P_{c} = 67.8\pm10.7\%$, $P_{xt}=13.4\pm19.4\%$, $N_0=8.53\pm4.3$ photons, and $P_a=52.3\pm26.8\%$.
The errors are large due to the degeneracy between having a free value of $P_a$ and $N$, which propagates through, and an independent measurement of either would reduce the errors considerably.
However, the errors reported here are utilized as systematic errors in this work as they are likely the same magnitude of applying the $P_c$ correction to different devices.

\end{document}